\documentclass[english,aps,prl,twocolumn,nopacs,groupedaddress,floatfix]{revtex4-1}

\usepackage{amsmath,amsfonts,amssymb,graphics,graphicx,epsfig,color,times,bbm}

\newcommand{\je}[1]{\textcolor{black}{#1}}

\begin{document}

\title{Opto- and electro-mechanical entanglement improved by modulation}

\author{A.\ Mari$^{1,2}$ and J.\ Eisert$^{1,2}$}

\affiliation{$^1$ Dahlem Center for Complex Quantum Systems, Freie Universit\"at
Berlin, 14195 Berlin, Germany}
\affiliation{$^2$ Institute of Physics and Astronomy, University of Potsdam,
D-14476 Potsdam, Germany}

\begin{abstract}
One of the main milestones in the study of opto- and electro-mechanical systems is to
certify entanglement between a mechanical resonator and an optical or microwave mode of
a cavity field. 
In this work, we show how a suitable time-periodic
modulation can help to achieve large degrees of entanglement, building upon the framework
introduced in [Phys.\ Rev.\ Lett.\ {\bf 103}, 213603 (2009)]. It is demonstrated that with suitable
driving, the maximum degree of entanglement can be significantly enhanced, in a way
exhibiting a non-trivial dependence on the specifics of the modulation.
Such time-dependent driving might help experimentally achieving entangled mechanical systems  
also in situations when quantum correlations are otherwise suppressed by thermal noise. 
\end{abstract}

\maketitle

\section{Introduction}
Opto-mechanical \cite{Opto1,Opto2,Opto3, Opto4, Opto5, KippenCold, AspelCold} 
and electro-mechanical systems \cite{EM0,EM1,EM2,EM3,EMCold, SB} 
are promising candidates for realizing architectures exhibiting quantum behavior in 
macroscopic structures. Once the quantum regime is reached, exciting applications
in quantum technologies such as realizing precise force sensors are conceivable \cite{Schwab, Marquardt}.
One of the requirements to render such an approach feasible, needless to say,
is to be able to certify that a mechanical degree of freedom is deeply in the quantum
regime \cite{Marquardt,QR1,QR2,QR3,QR4}. The detection of 
entanglement arguably constitutes the ultimate benchmark in this respect. 
While effective ground state cooling has indeed been experimentally closely approached 
\cite{EMCold,KippenCold} and achieved \cite{SB, EM0, AspelCold}, 
the detection of entanglement is still awaiting.

In this work, we emphasize that a mere suitable time-modulation of the 
driving field may significantly help to achieve entanglement between a mechanical 
mode and a radiation mode of the system.
We extend the idea of Ref.\ \cite{Ours}, putting emphasis on the improvement 
of entanglement by means of suitable modulations \cite{Ours,Clerk,Woolley}. 
The method used here is not a direct modulation of the frequencies of the two modes
(parametric amplification), but the system is instead externally driven with a modulated 
field. This time dependence of the driving indirectly affects the effective radiation pressure 
coupling between the two modes and generates non-trivial entanglement resonances. 
In this way, with the appropriate choice of the modulation pattern, large degrees 
of two-mode squeezing can be reached.

\section{Modulated opto- and electro-mechanical systems}

We consider the simplest scenario of a mechanical resonator of frequency
$\omega_m$ coupled to a single mode of the electromagnetic field of frequency
$\omega_a$. This radiation field could be an optical mode of a Fabry-Perot
cavity \cite{Opto1,Opto2,Opto3,Opto4,Opto5, SidebandCooling, KippenCold,AspelCold,QR2,QR3} 
or a microwave mode of a superconductive circuit
\cite{EM0, EM3, EMCold, EMtheory}.
It can be shown that the Hamiltonians associated to this two
experimental settings are formally equivalent \cite{QR3,EMtheory} and therefore the theory that we
are going to introduce is general enough to describe both types of
systems. 

We assume that the radiation mode is driven by a coherent field with a time
dependent amplitude $E(t)$ and frequency $\omega_l$. The particular choice of
the time dependence is left unspecified but we impose the structure of a
periodic modulation such that $E(t+\tau)=E(t)$ for some $\tau>0$ of the order
of $\omega_m^{-1}$. In this sense, the driving regime that we are going to study is 
intermediate between the two opposite extremes of constant amplitude and short pulses.
The Hamiltonian of the system is
\begin{eqnarray}
	&& H=\hbar \omega_a
a^{\dagger}a+\frac{1}{2}\hbar\omega_{m}(p^{2}+q^{2})-
	\hbar g a^{\dagger}a q \nonumber  \\ 
	&& +i\hbar  [ E(t)e^{-i \omega_l t}a^{\dagger}- E^*(t)e^{i \omega_l t}
a],
	\label{ham0}
\end{eqnarray}
where the mechanical mode is described in terms of dimensionless position and
momentum operators satisfying
$[q,p]=i$, while the radiation mode is captured by creation and annihilation
operators obeying the bosonic commutation rule $[a,a^\dag]=1$. The two modes
interact via a radiation pressure potential with a strength given by the
coupling parameter $g$.

In addition to this coherent dynamics, the mechanical mode will be unavoidably
damped at a rate $\gamma_m$, while the optical/microwave mode will decay at a
rate $\kappa$. 
These dissipative processes and the associated fluctuations can be taken into
account in the Heisenberg
picture by the following set of quantum Langevin equations \cite{QR1,QR2,QR3, EMtheory},
\begin{eqnarray} \label{QLE}
\dot{q}&=&\omega_m p , \\
\dot{p}&=&-\omega_m q - \gamma_m p + g a^{\dag}a + \xi,  \nonumber \\
\dot{a}&=&-(\kappa+i\Delta)a +i g a q +E(t) +\sqrt{2\kappa} a^{\rm in}. \nonumber
\end{eqnarray}
In this set of equations a convenient rotating frame has be chosen $a \mapsto a e^{-i \omega_l
t}$, such that
the detuning parameter is $\Delta=\omega_a-\omega_l$.
The operators $\xi$ and $a^{\rm in}$ represent the mechanical and optical bath
operators respectively,
and their correlation functions are well approximated by delta functions
\begin{eqnarray} \label{Marknoise}
	 \langle \xi(t) \xi (t')+ \xi(t') \xi (t) \rangle/2&=&\gamma_m (2 n_m
+1) \delta(t-t'), \\
          \langle a^{\rm in}(t) a^{\rm in \dag}(t') \rangle &=& (n_a+1) \delta(t-t'),
\nonumber \\
          \langle a^{\rm in \dag}(t) a^{\rm in}(t') \rangle &=&n_a\delta(t-t')
,\nonumber
\end{eqnarray}
where $n_x=(\text{exp}{( {\hbar \omega_x}/({k_B T}))}-1)^{-1}$, is the bosonic
mean occupation number
at temperature $T$.

\section{Classical periodic orbits: first moments}

We are interested in the coherent strong driving regime when $\langle a \rangle \gg1$. In
this limit, 
the semiclassical approximations $\langle a^{\dag}a \rangle\simeq |\langle a
\rangle|^2$ and
$\langle a q \rangle \simeq \langle a \rangle \langle q \rangle$ are good approximations.
Within this approximation, one can average
both sides of Eq.\ (\ref{QLE}) and get a differential equation for the
first moments of the canonical coordinates
\begin{eqnarray}\label{QLEav}
 \langle \dot{q}  \rangle &=&\omega_m  \langle p  \rangle , \\
 \langle \dot{p}  \rangle &=&-\omega_m  \langle q  \rangle - \gamma_m  \langle p
 \rangle + g  |\langle a \rangle|^2 ,
 \nonumber  \\
 \langle \dot{a} \rangle &=&-(\kappa+i\Delta) \langle a  \rangle +i g  \langle a
 \rangle  \langle q  \rangle+E(t).\nonumber
\end{eqnarray}

Far away from the well known opto- and electro-mechanical instabilities, asymptotic
$\tau$-periodic solutions can be used as ansatz for Eqs.\ (\ref{QLEav})   (see the Appendix 
for a more detailed analysis). 
These solutions represent periodic orbits in phase space and are usually called limit cycles. 
These cycles are induced by the modulation and
should not be confused with the limit cycles emerging in the strong driving
regime due to the non-linearity of the system. Because of the asymptotic
periodicity of the solutions, one can define the fundamental modulation frequency
as $\Omega=2 \pi / \tau$, such that each periodic solution can be expanded
in the following Fourier series
\begin{eqnarray}\label{fourier}
 \langle O (t) \rangle  &=& \sum_{n=-\infty}^{\infty} O_n e^{i n \Omega t},
\quad O=q,p,a. 
\end{eqnarray}
The Fourier coefficients $\{O_n\}$ appearing in Eq.\ (\ref{fourier}) can be
analytically estimated as shown in Appendix
and they completely characterize the classical asymptotic dynamics of the system.

Finally we notice that the classical evolution of the dynamical variables will
shift the detuning to the effective value of $\tilde \Delta(t)=\Delta-g \langle
q (t) \rangle$. For the same reason, it is also convenient to introduce an
effective coupling constant defined as 
\begin{equation}	
	\tilde g(t)=i g \langle a(t)
	\rangle/\sqrt{2}.
\end{equation}

\section{Quantum correlations: second moments}

The classical limit cycles are given by the asymptotic solutions of Eqs.\
(\ref{QLEav}).
In order to capture the quantum fluctuations around the classical orbits, we
introduce a column vector of new quadrature operators $u=[\delta q,\delta
p,\delta x,\delta y]^T$ defined as:
\begin{eqnarray}
 \delta q &=& q -\langle q(t) \rangle  , \label{flu}\\
 \delta p &=& p -\langle p(t) \rangle  , \nonumber \\
 \delta x &=& \left[ (a -\langle a(t) \rangle)+ (a -\langle a(t) \rangle)^\dag
\right] / \sqrt{2} , \nonumber\\
 \delta y &=&- i \left[ (a -\langle a(t) \rangle)- (a -\langle a(t) \rangle)^\dag
\right] / \sqrt{2}. \nonumber
 \end{eqnarray}
This set of canonical coordinates 
can be viewed as describing a time-dependent reference frame co-moving with the
classical orbits. 
The corresponding vector of noise operators will be 
\begin{equation}
n=[0,\xi ,(a^{\rm in}+a^{\rm in \dag})/\sqrt{2},-i(a^{\rm in}-a^{\rm in \dag})/\sqrt{2} ]^T.
\end{equation}
Since we are in the limit in which classical orbits emerge ($\langle a \rangle
\gg1$), it is a reasonable approximation to
express the previous set of Langevin equations (\ref{QLE}) in terms of the new
fluctuation operators (\ref{flu})
and neglect all quadratic powers of them. The resulting linearized system can be
written as a matrix equation \cite{Ours},
\begin{eqnarray}\label{LQLE}
\dot u= A(t) u+n(t) ,
\end{eqnarray}
where,
\begin{eqnarray}
  A(t)&=&\left[\begin{array}{cccc}
    	0 	  & \omega_m	& 0 		& 	0 \\
     	-\omega_m & -\gamma_m	& \Re \tilde g(t)	& \Im \tilde g(t)  \\
        - \Im \tilde g(t)	  & 	0	& -\kappa 	& \tilde
\Delta(t) \\
    	 \Re \tilde g(t)   &     0	& -\tilde \Delta(t) 	& -\kappa
  	\end{array}\right]
\end{eqnarray}
is a real time-dependent matrix.

If the system is stable, and as long as the linearization is valid, the quantum
state of the system will converge 
to a Gaussian state with time dependent first and second moments. The first
moments of the state correspond to the 
classical limit cycles introduced in the previous section. The second moments
can be expressed in terms of
the covariance matrix $V(t)$ with entries
\begin{equation}
 V_{k,l}(t)=\langle u_k(t) u_l^\dag(t)+u_l^\dag(t) u_k(t)\rangle/2.
\end{equation}
One can also define a diffusion matrix $D$ as
 \begin{equation} \label{Dnoise}
 \delta (t-t')D_{k,l}=\langle n_k(t) n_l^\dag(t')+n_l^\dag(t') n_k(t)\rangle/2,
\end{equation}
which, from the properties of the bath 
operators (\ref{Marknoise}), is diagonal and equal to 
\begin{equation}
D={\rm diag}[0,\gamma(2 n_m+1),\kappa (2 n_a+1),\kappa (2 n_a+1)].
\end{equation}
From Eqs.\ (\ref{LQLE}) and (\ref{Dnoise}), one can easily derive a linear differential equation for
the correlation matrix,
 \begin{equation}
 \frac{d}{dt}V(t)=A(t)V(t)+V(t)A^T(t)+D. \label{Vdot}
\end{equation}
Since the first and the second moments are specified,
Eqs.\ (\ref{QLEav}) and (\ref{Vdot}) provide a complete 
description of the asymptotic dynamics of the system.  Apart
from the linearization around classical cycles, no further approximation has been
done: Neither a weak coupling, adiabatic or rotating-wave approximation.
Numerical solutions of both equations (\ref{QLEav}) and (\ref{Vdot}) can be
straightforwardly found. These solutions will be used to calculate the exact
amount opto- and electro-mechanical entanglement present in the system.

The asymptotic periodicity of the classical solutions (Eq.\ (\ref{fourier}))
implies that, in the long time limit, $A(t+\tau)=A(t)$. This means that Eq.
(\ref{Vdot}) is a linear differential equation with periodic coefficients and
then all the machinery of Floquet theory is in principle applicable. Here,
however, since we are only interested on asymptotic solutions, we are not going
to study all the Floquet exponents of the system. The only property that we need
is that, in the long time limit, stable solutions will acquire the same
periodicity of the coefficients: 
\begin{equation}
	V(t+\tau)=V(t). 
\end{equation}
This is a simple corollary of Floquet's theorem. 
In the subsequent sections we will apply the previous theory to some particular
experimental setting and show how a simple 
modulation of the driving field can significantly improve the amount of
opto- and electro-mechanical entanglement. 

\section{Entanglement resonances}

In this section we are going to study what kind of amplitude modulation is optimal
for generating entanglement between the radiation and mechanical modes. 
As a measure of entanglement we use the logarithmic negativity $E_N$ which, since the
state is Gaussian, can be easily computed directly from the correlation
matrix $V(t)$ \cite{Neg1,Neg2,Neg3}.
We have also seen that the correlation matrix is, in the long time limit,
$\tau$-periodic.
This suggests that it is sufficient to study the variation of entanglement in a
finite interval of time $[t,t+\tau]$ for large times $t$.
One can then define the maximum amount of achievable entanglement as 
\begin{equation}
	\hat
	E_{N}=\lim_{t\rightarrow \infty}\max_{h \in [t,t+\tau]} E_N (h). 
\end{equation}
This will be
the quantity that we are going to optimize.

\begin{figure}[tbh]
\centerline{\includegraphics[width=0.4\textwidth]{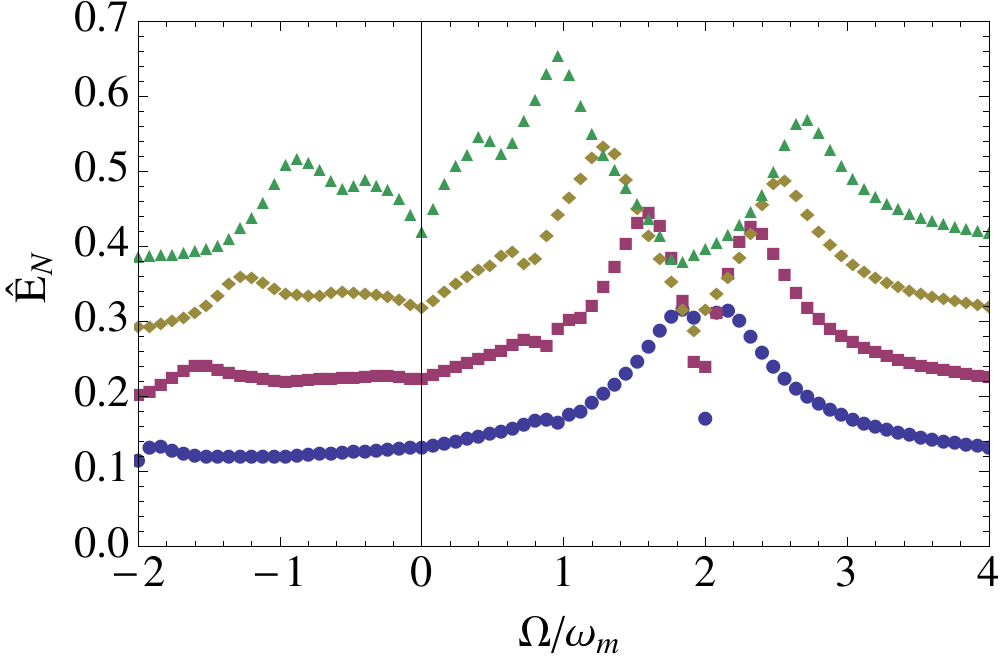}} 
\caption{Maximum entanglement $\hat E_N$ as a function of the modulation
frequency $\Omega$ and for different values of the driving strength $\tilde g_0$. 
The chosen parameters in units of $\omega_m$ are:
$\kappa=0.2$, 
$\gamma_m=10^{-6}$, 
$\tilde \Delta=1$, 
$n_m=n_a=0$,
$\tilde g_\Omega=0.1$, 
$\tilde g_0=$
$0.2$ (circles),
$0.4$ (squares),
$0.6$ (diamonds),
$0.8$ (triangles).
} \label{fig-omega}
\end{figure}

We first study a very simple set of parameters (see caption of Figure 
\ref{fig-omega}) in order to understand what the optimal choice is
for the modulation frequency. For this purpose we impose the effective coupling
to have this simple structure
\begin{equation}\label{gt}
\tilde g(t)= \tilde g_0+ \tilde g_{\Omega} \; e^{-i \Omega t},
\end{equation}
where $\tilde g_0$ is associated to the main driving field with detuning
$\Delta$, while $\tilde g_\Omega$ is the amplitude of a further sideband shifted
by a frequency $\Omega$ from the main carrier. Without loss of generality
we will assume $\tilde g_0$ and $\tilde g_\Omega$ to be positive reals. 
This kind of driving is a natural one and has been chosen for reasons
that will become clear later.
From now on we set the detuning of the carrier frequency
to be equal to the mechanical frequency $\Delta=\omega_m$. This choice of the
detuning corresponds to the well known sideband cooling setting \cite{SidebandCooling, QR1} and
it has been shown to be also optimal for maximizing opto-mechanical entanglement
with a non-modulated driving \cite{QR3}. 
Fig.\ \ref{fig-omega} shows the maximum entanglement $\hat E_N$ between the
mechanical and the radiation modes as a function of the modulation frequency
$\Omega$ and for different values of the driving amplitude $\tilde g_0$. This maximum degree of 
entanglement has been calculated for $t>200/\kappa$ when the system
has well reached its periodic steady state.

We observe that in Fig.\ \ref{fig-omega} there are two main resonant peaks at the modulation frequencies
\begin{equation}\label{omegapm}
	\Omega\simeq 2\omega_m \pm \tilde g_0.
\end{equation} 
\je{We will now provide some intuition why one should expect the main resonances at the 
locations where they are observed. First assume that $\tilde g_0=0$. Then, for  $\Delta=\omega_m$, the linearized Hamiltonian in the interaction picture is 
\begin{eqnarray}
	H_{\rm int}&=&-\hbar \tilde g_\Omega \left( e^{i (\omega_m-\Omega)}\delta a^\dag 
	+ e^{-i (\omega_m-\Omega)}\delta a  \right)
	\nonumber \\
	&&\left( e^{i \omega_m} \delta b^\dag +\delta b  e^{-i \omega_m} \right)/2, \label{Hint}
\end{eqnarray}
where the bosonic operators are defined as $\delta a=(\delta x+i \delta y)/\sqrt{2}$, $\delta b=(\delta q+i\delta p)/\sqrt{2}$.
From Eq.\ (\ref{Hint}), it is clear that for $\Omega=2 \omega_m$, neglecting all rotating terms, we get the well known two-mode squeezing generator
\begin{eqnarray}\label{tms}
	H_{\rm int}&\simeq&-\hbar \tilde g_\Omega \left( \delta a^\dag \delta b^\dag + \delta a \delta b \right) /2.
\end{eqnarray}
So, in the case of $\tilde g_0=0$, a modulation of $\Omega=2 \omega_m$ would be the most
reasonable choice in order to generate entanglement. 
However, this regime is well known to be highly unstable and, in practice, it cannot be used for preparing entangled
steady states \cite{EMtheory}.}

This is why we need to consider a modulated coupling of the form given in Eq.\ (\ref{gt}) -- or a similar type of
modulation sharing these features. We now allow for $\tilde g_0$ being different from zero, giving rise to a
situation which can be assessed
in a very similar way as above (only that the rotation terms will take a more involved form). 
The main amplitude $\tilde g_0$ then takes the role of cooling and stabilizing the system while the modulation amplitude $\tilde g_\Omega$ is used to generate entanglement. At the same time however, as shown in Refs.\ \cite{ModeSplitting, Opto4}, for  
$\tilde g_0 > \kappa/\sqrt{2}$ the system hybridizes in two normal modes of frequencies 
\begin{equation}
	\omega_{\pm}\simeq \omega_m \pm \tilde g_0/2.
\end{equation}	
	As a consequence, this will affect the modulation frequency $\Omega$ that one 
	has to choose in order to achieve the two-mode squeezing interaction given in Eq. (\ref{tms}). This is the reason for the presence of two resonant peaks in Fig.\ \ref{fig-omega} and for the resonance condition given in Eq.\ (\ref{omegapm}).

Note also that the choices of modulations that give rise to the optimal local single-mode squeezing \cite{Ours}
of the mechanical mode and the degree of entanglement are not identical. This is rooted in the ``monogamous
nature'' of squeezing: For a fixed spectrum of the covariance matrix, one can either have large local or
two-mode squeezing. This effect is observed when considering the modulation 
frequencies that achieve
maximum single- and two-mode squeezing.

\begin{figure}[tbh]
\centerline{\includegraphics[width=0.4\textwidth]{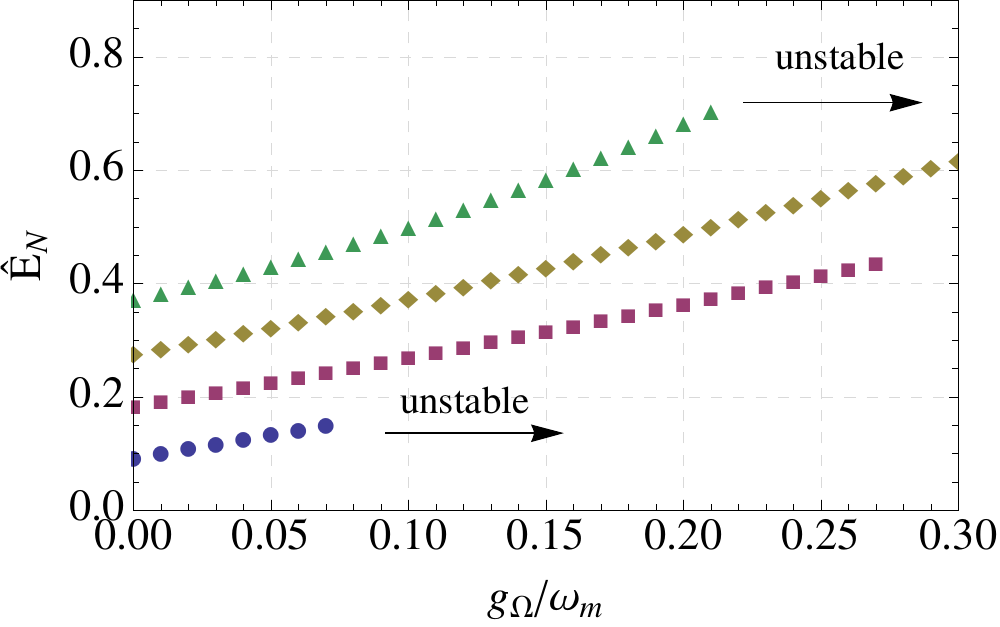}} 
\caption{Maximum entanglement $\hat E_N$ as a function of the modulation
amplitude $\tilde g_\Omega$ and for different values of the driving strength $\tilde g_0$. 
The chosen parameters in units of $\omega_m$ are:
$\kappa=0.2, \gamma_m=10^{-6}$, 
$\tilde \Delta=1$, 
$n_m=n_a=0$,
$\Omega=2\omega_m-\tilde g_0$, 
$\tilde g_0=0.2$ (circles),
$0.4$ (squares),
$0.6$ (diamonds),
$0.8$ (triangles).
}  \label{fig-gg}
\end{figure}

We finally observe that the height of the two peaks, due to the cavity filtering, is not equal: the first 
resonance at $\Omega= 2\omega_m - \tilde g_0$ is better for the amount of steady state entanglement.
One could also ask what
the behavior of entanglement is when we change the amplitude of the modulation.
Fig.\ \ref{fig-gg} shows the amount of entanglement $\hat E_N$ as a
function of $\tilde g_\Omega$ and for different choices of $\tilde g_0$. We observe
that entanglement is monotonically increasing in $\tilde g_\Omega$ up to a threshold 
where the system becomes unstable.

\section{Opto- and electro-mechanical entanglement in realistic settings}

\begin{figure}[tbh]
\leftline{\includegraphics[width=0.5\textwidth]{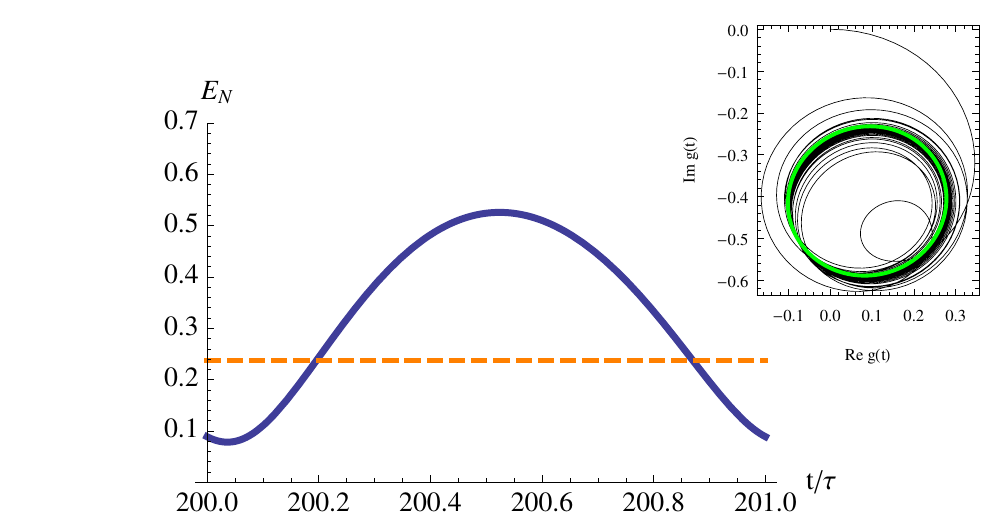}} 
\caption{(Optical cavity). The degree of entanglement, measured in terms of the
 logarithmic negativity, as a function of time. The
full line refers to a modulated driving ($\Omega=1.4 \omega_m$) while the
dotted line corresponds to a non-modulated driving ($\Omega=0$).
The chosen parameters in units of $\omega_m$ are:
$\kappa=0.2$, 
$\gamma_m=10^{-6}$,
$\Delta=1$, 
$n_m=2\times 10^3$, 
$n_a=0$,
$g_0=4\times10^{-6}$, 
$E_0=7\times10^4$,
$E_\Omega=2.5\times10^4$. 
The inset shows the trajectory of the effective coupling $\tilde g(t)=\sqrt{2} g \langle a(t) \rangle$
in the complex plane due to the time evolution of the optical amplitude. The phase space orbit (black line) is 
numerically simulated from Eq.\ (\ref{QLEav}), while the limit cycle (green line) is an analytical
approximation (see Appendix for more details).} 
\label{fig-optics}
\end{figure}

\begin{figure}[tbh]
\centerline{\includegraphics[width=0.5\textwidth]{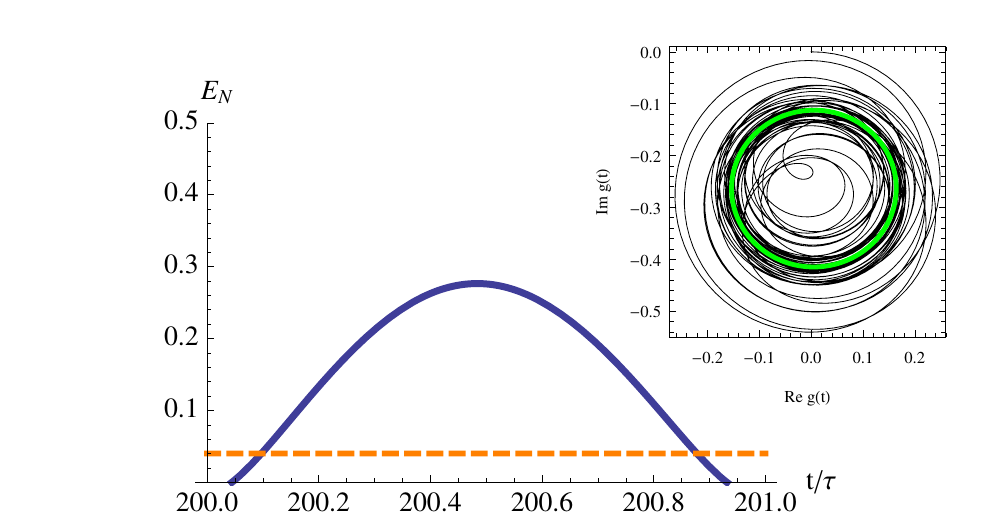}} 
\caption{(Microwave cavity). Entanglement log-negativity as a function of time. The
full line refers to a modulated driving ($\Omega=1.3 \omega_m$) while the
dotted line corresponds to a non-modulated driving ($\Omega=0$).
The chosen parameters in units of $\omega_m$ are:
$\kappa=0.02$, 
$\gamma_m=3\times 10^{-6}$, 
$\Delta=1$, 
$n_m=200$, 
$n_a=0.03$,
$g_0=2\times10^{-5}$, 
$E_0=9\times10^3$,
$E_\Omega=1.3\times10^3$. 
The inset depicts the trajectory of the effective coupling $\tilde g(t)=\sqrt{2} g \langle a(t) \rangle$
in the complex plane due to the time evolution of the microwave amplitude. The phase space orbit (black line) is 
numerically simulated from Eq.\ (\ref{QLEav}), while the limit cycle (green line) is an analytical
approximation (see Appendix for more details).}  \label{fig-micro}
\end{figure}

We have seen that an effective coupling of the form $\tilde g(t)=\tilde
g_0+\tilde g_{\Omega} \; e^{-i (2\omega_m -\tilde g_0)t}$ is optimal for the 
generation 
of entanglement within the considered class of drivings. 
However, the parameter $\tilde g(t)$ depends on the average 
amplitude $\langle a(t) \rangle$ and assuming such a simple structure may
seem somewhat artificial.
In this section, we show how the desired time-dependent coupling can indirectly result
from the classical limit cycles of
the system (see insets of Figs.\ \ref{fig-optics} and \ref{fig-micro}) and we also take into account the
 effect of a temperature of the order of $T\simeq 100$ mK.  
The natural ``educated guess'' for the structure of the driving field will be
\begin{equation}
	E(t)=E_0+E_\Omega E e^{-i (2 \omega_m -\tilde g_0)t}.
\end{equation}	 
For the choice of the other parameters,
we focus on two set of parameters corresponding to two completely
different systems:
an optical cavity with a moving mirror and a superconducting wave guide coupled
to a mechanical resonator. The parameters are chosen according to realistic
experimental settings, see, e.g., Ref.\ \cite{Opto4} (opto-mechanical system) and
Ref.\ \cite{EM0} (electro-mechanical system). Fig.\ \ref{fig-optics} and
Fig.\ \ref{fig-micro} show that, in both experimental scenarios, entanglement
can significantly be increased by an appropriate modulation of the driving
field.

\section{Summary}

In this work, we have shown how time-modulation can significantly enhance the 
maximum degree of entanglement. Triggered by the time-modulated driving, 
the mode of the electromechanical field as well as the mechanical mode start ``rotating around
each other'' in a complex fashion, giving rise to increased
degrees of entanglement. The dependence on the frequencies of the additional modulation is intricate, with
resonances highly improving the amount of entanglement that can be reached. The ideas presented
here could be particularly beneficial to prepare systems in entangled states
in the first place, in scenarios where the parameters are such that the states prepared are
close to the boundary to entangled states, but where this boundary is otherwise not yet
quite reachable with present technology. At the same time, such ideas are expected to 
be useful in metrological applications whenever high degrees of entanglement are needed.

\section{Appendix}

In this appendix we derive analytical formulas for the asymptotic solutions of
the classical system of dynamical equations
(\ref{QLEav}). 
A crucial assumption for the following procedure is that it is possible to
expand the solutions in powers of the the coupling
constant $g_0$
\begin{equation}\label{G0expansion}
	\langle O \rangle(t)=\sum_{j=0}^{\infty} O_j(t) g_0^j, \end{equation}
where $O=a,p,q$. 
This is justified only if the system is far away from multi-stabilities and the
radiation pressure coupling can be treated in a perturbative way. A very
important feature of the set of equations (\ref{QLEav}) is that they contain
only two non linear terms and those terms  are proportional to the coupling
parameter $g_0$. This implies that, if we use the ansatz (\ref{G0expansion}),
each function $O_j$ will be a solution of {\it linear} differential equation
with time dependent parameters depending on the previous solution $O_{j-1}(t)$.
Since $E(t)=E(t+\tau)$, from a recursive application of Floquet's theorem,
follows that stable solutions will converge to periodic limit cycles having the
same periodicity of the driving: $\langle O(t) \rangle=\langle O(t+\tau)
\rangle$.
One can exploit this property and perform a double expansion in powers of $g_0$
and in terms of Fourier components
\begin{equation}\label{double}
\langle O \rangle(t)= \sum_{j=0}^{\infty}\sum_{n=-\infty}^{\infty} O_{n,j} e^{i
n \Omega t}g_0^j,
\end{equation}
where $n$ are integers and $\Omega=2\pi /\tau$.
A similar Fourier series can be written for the periodic driving field,
\begin{equation}
E (t)= \sum_{n=-\infty}^{\infty} E_n e^{i
n \Omega t}.    
\end{equation}
The coefficients $O_{n,j}$ can be found by direct substitution in Eq.\
(\ref{QLEav}). They are completely determined by the following set of recursive
relations:
\begin{equation}
 	q_{n,0}=p_{n,0}=0, \qquad a_{n,0}=\frac{E_{-n}}{\kappa	
	+i(\Delta_0 + n \Omega)},
\end{equation}
corresponding to the $0$-order perturbation with respect to $G_0$, and 
\begin{eqnarray}
	p_{n,j}&=&\frac{i n \Omega}{\omega_m} q_{n,j}, \\
	q_{n,j}&=&\omega_m \sum_{k=0}^{j-1}\sum_{m=-\infty}^{\infty} 
	\frac{  a_{m,k}^*\; a_{n+m,j-k-1}}{ \omega_m^2-n
\Omega^2+i\gamma_m n \Omega}, \\
	a_{n,j}&=&i \sum_{k=0}^{j-1}\sum_{m=-\infty}^{\infty} \frac{a_{m,k} 
	q_{n-m,j-k-1}}{\kappa +i (\Delta_0 + n \Omega)},
\end{eqnarray} 
giving all the $j$-order coefficients in a recursive way.
For all the examples analyzed in this paper we truncated the analytical
solutions up to $j\le 3$ and $|n|\le2$. This level of approximation is already
high enough to well reproduce the exact numerical solutions.

\section{Acknowledgements}
	We would like to thank the EU (MINOS, COMPAS, QESSENCE) and the BMBF (QuOReP)
	for support.

\end{document}